\documentclass[twocolumn]{aastex62}
\usepackage{longtable}
\usepackage{amsmath}
\usepackage{subfigure}
\usepackage{lipsum}  

\usepackage[normalem]{ulem}

\DeclareMathOperator{\sech}{sech}
\DeclareMathOperator{\arcsech}{arcsech}
\DeclareMathOperator{\arcsinh}{arcsinh}

\graphicspath{{./}{figures/}}

\usepackage{scrextend}

\begin{document}

\title{Impact of Ejecta Temperature and Mass on the Strength of Heavy Element Signatures in Kilonovae}

\author[0000-0002-9852-2469]{Donggeun Tak}
\affiliation{SNU Astronomy Research Center, Seoul National University, Seoul 08826, Republic of Korea, \href{donggeun.tak@gmail.com}{donggeun.tak@gmail.com}}
\affiliation{Korea Astronomy and Space Science Institute, Daejeon 34055, Republic of Korea}
\author{Z. Lucas Uhm}
\affiliation{Korea Astronomy and Space Science Institute, Daejeon 34055, Republic of Korea}
\author[0000-0002-8094-6108]{James H. Gillanders}
\affiliation{Astrophysics sub-Department, Department of Physics, University of Oxford, Keble Road, Oxford, OX1 3RH, UK}

\correspondingauthor{Donggeun Tak (\href{donggeun.tak@gmail.com}{donggeun.tak@gmail.com})}

\begin{abstract}

A kilonova, the electromagnetic emission produced by compact binary mergers, is formed through a delicate interplay of physical processes, involving r-process nucleosynthesis and interactions between heavy elements and photons through radiative transfer. This complexity makes it difficult to achieve a comprehensive understanding of kilonova spectra. In this study, we aim to enhance our understanding and establish connections between physical parameters and observables through radiative-transfer simulations. Specifically, we investigate how ejecta temperature and element mass influence the resulting kilonova spectrum. For each species, the strength of its line features depends on these parameters, leading to the formation of a distinct region in the parameter space, dubbed the Resonance Island, where the line signature of that species is notably evident in the kilonova spectrum. We explore its origin and applications. Among explored r-process elements (31$\leq$Z$\leq$92), we find that four species -- Sr\textsubscript{II}, Y\textsubscript{II}, Ba\textsubscript{II}, and Ce\textsubscript{II} -- exhibit large and strong resonance islands, suggesting their significant contributions to kilonova spectra at specific wavelengths. In addition, we discuss potential challenges and future perspectives in observable heavy elements and their masses in the context of the resonance island.
\end{abstract}

\keywords{Atomic data, Line: identification, Stars: neutron, radiative transfer, Astrophysics -- High Energy Astrophysical Phenomena, Astrophysics -- Solar and Stellar Astrophysics}

\section{Introduction}

The kilonova spectrum emerges as a consequence of intricate physical processes and interactions associated with the merging of two neutron stars or a neutron star and a black hole, along with subsequent outflows, including dynamical and disk-wind ejecta (see \citealt{Metzger2019} and \citealt{Barnes2020} for the recent review of kilonovae). The merging process initiates r-process nucleosynthesis, followed by consequential interactions between the produced heavy elements and a flood of photons resulting from radioactive decay. Within the ejecta, these interactions give rise to distinct spectral lines in the kilonova spectrum. The complexity inherent in these processes poses a challenge, making it difficult to standardize or establish a uniform pattern for the kilonova spectrum. 

While understanding the entire kilonova process with sophisticated yet computationally intensive magneto-hydrodynamic simulations of a binary merger is demanding, the cause-and-effect relationships within the radiative-transfer process offer a potentially more accessible path for investigation. Although not fully explored, this approach also holds significant potential for shedding light on kilonova physics. For example, vigorous studies on AT2017gfo \citep[e.g.,][]{Kasen2017, Perego2017, Siegel2017, Tanaka2017, Radice2018, Ford2024, Vieira2024} highlighted the key role of $Y_e$ in shaping the spectrum, with higher $Y_e$ generally leading to bluer spectra. However, the full parameter space of $Y_e$ and its impact remains largely unexplored. 

The kilonova spectrum from the radiative transfer process is shaped by the complicated interplay of ejecta properties, such as electron fraction ($Y_e$), velocity and density profiles, and temperature. These parameterized ejecta configurations have been widely studied \citep[e.g.,][]{Tanaka2013, Kasen2017, Even2020, Bulla2021, Wu2022, Gillanders2022, Gillanders2023, Fontes2023, Shingles2023, Sneppen2023a, Tak2023, Fryer2024}. In \cite{Tak2023}, we delved into the influence of diverse ejection histories on the development of kilonova spectra and their associated light curves. Specifically, we introduced representative velocity profiles of ejecta and explored how features in each velocity profile leave their imprints on both the kilonova spectrum and the light curve. \cite{Fryer2024} also conducted a study with a similar focus, aiming to understand the impact of velocity distribution on light curves and spectra. As discussed by \cite{Shingles2023} and \cite{Collins2023}, line transitions by Sr\textsubscript{II} commonly occur at the front end of the expanding line-forming region with higher velocities. This indicates a connection between the strength of line transitions and the velocity profile of the ejecta. The role of ejecta mass and its velocity was also emphasized by \cite{Fontes2023}.

Other than the ejecta mass and velocity, various properties of kilonovae have been widely studied. For example, the influence of different compositions and their opacities on the shape of a kilonova spectrum has been widely investigated \citep[e.g.,][]{Even2020, Tanaka2020, Fontes2023}. \cite{Korobkin2021} examined various morphologies of kilonovae and their impact on the light curve. Additionally, \cite{Sneppen2023a} discussed the relationship between the electron density in the ejecta and the strength of the Sr\textsubscript{II} line feature.

In conjunction with these efforts, in this study, our goal is to illustrate how the strength and occurrence of line transitions change with the ejecta temperature and element mass. Details on the parameterized configurations of the ejecta model are provided in Section~\ref{sec:setup}. We perform a radiative-transfer spectral analysis for the specified parameter spaces, and the results are presented in Section~\ref{sec:result}. In Section~\ref{sec:disc}, we discuss the results and their implications, while Section~\ref{sec:summary} provides a summary of our findings.

\section{Model Configuration}\label{sec:setup}

\begin{table*}
    \centering
    \caption{
        Solar r-process abundance for 11 selected elements, adopted from \cite{Prantzos2020}. The selection is based on their contributions to the spectrum, where the listed elements contribute more than 3\% of the total bolometric luminosity at some temperature and ejecta mass (see Section~\ref{sec:map} and Figure~\ref{fig:ri_map}). Note that we re-normalize the mass fraction with respect to the mass of Sr by setting the mass fraction of Sr to 1. Lanthanides are elements with atomic numbers, $58 \leq Z \leq 71$. 
    }
    \begin{tabular}{l| c*{10}{c}}
        \hline
        \textbf{Element} & Sr & Y & Zr & Ba & La & Ce & Pr & Nd & Tb & Dy & Ho\\\hline
        \textbf{Atomic number (Z)} & 38 & 39 & 40 & 56 & 57 & 58 & 59 & 60 & 65 & 66 & 67\\
        \textbf{Mass fraction} & 1.00 &0.54 &1.06 &0.39 &0.08 &0.14 &0.07 &0.28 &0.05 &0.33 &0.08 \\\hline
    \end{tabular}
    \label{tab:abud}
\end{table*}

We conduct radiative transfer spectral analysis across diverse kilonova ejecta conditions using \textsc{tardis} \citep[][]{Kerzendorf2014, tardis}. This 1D Monte Carlo radiative transfer spectral synthesis code has been employed extensively to simulate kilonova spectra \citep[e.g.,][]{Smartt2017, Watson2019, Gillanders2021, Gillanders2022, Perego2022, Tak2023}.
The basic mechanisms for how \textsc{tardis} generates synthetic spectra are summarised below, but for full details, see \cite{Kerzendorf2014} or the online documentation.\footnote{\url{https://tardis-sn.github.io/tardis/index.html}}

First, \textsc{tardis} defines a completely opaque, optically thick inner boundary, that emits a single-temperature blackbody. This opaque inner region is surrounded by an envelope of less opaque material, discretised into shells. These shells represent the computational domain, or line-forming region, of the simulation.\footnote{The model properties of the inner boundary and the envelope are controlled by user inputs.}
The code initializes a (user-defined) number of radiation packets, or $r$-packets (analogous to bundles of photons), at the surface of this photosphere (i.e.\ the inner boundary of the computational domain). 

The simulation begins with the $r$-packets randomly assigned both a propagation direction, and a frequency sampled from a (single-temperature) blackbody distribution, and
they freely stream through the computational volume until they interact. These $r$-packets can interact either via free-electron scattering, or some line interaction with a species (an atom or ion) present in the ejecta. When an interaction is triggered, a new propagation direction is assigned, and the $r$-packet continues along its path.

The propagation of these $r$-packets continues until they have all either escaped the outer boundary of the envelope (i.e.\ the outer boundary of the computational domain) or scattered beneath the photosphere. The properties of the escaping $r$-packets are then used to compute a synthetic spectrum. Each $r$-packet can undergo many interactions, depending on the specific model properties, but only the final interaction contributes to the synthetic spectrum.\footnote{While this statement is true for the real packet case, there exists a virtual packet treatment within \textsc{tardis} that can be used to increase the emergent packet count, effectively decreasing the noise introduced from Monte Carlo processes. This treatment depends on the intermediate interactions that each of the $r$-packets undergo \citep[see][for details]{Kerzendorf2014}.}

By tracking the interaction history of $r$-packets, \textsc{tardis} is capable of estimating the effective contributions of different species present in the line-forming region of the simulation to the emergent synthetic spectrum. The final synthetic spectrum will be composed of contributions from emergent $r$-packets that, in the line-forming region, either (1) did not undergo any interaction, (2) scattered with a free electron, or (3) underwent a line interaction with some species.

The properties of the ejecta material within the line-forming region (e.g.\ density, temperature, composition) are all governed by user-defined inputs. The following describes the model configurations for this simulation study.

\begin{figure}
  \centering
  \includegraphics[width=0.48\textwidth]{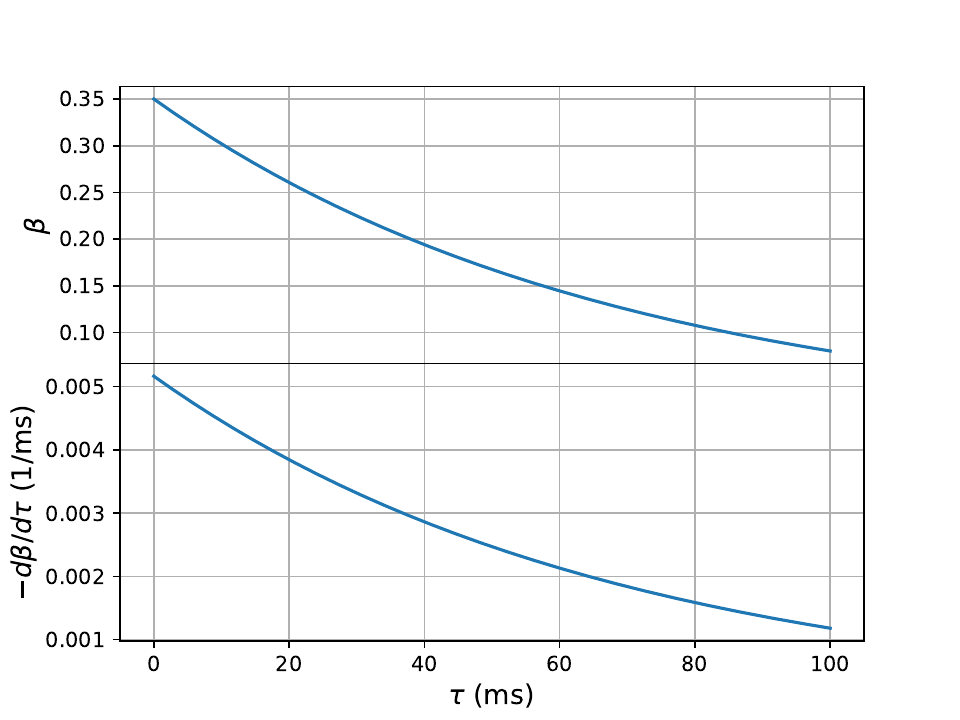}
  \caption{Ejecta velocity and its derivative profiles. The upper panel displays the velocity profile (Equation~\ref{eq:v_prof}), while the lower panel illustrates its derivative. Mass ejection is assumed to continue for up to 100 ms, with minimum and maximum velocities of 0.08~c and 0.35~c, respectively.}
  \label{fig:velocity}
\end{figure}

First, the velocity profile of the ejecta, represented as a function of ejection time $\tau$, is assumed to be
\begin{equation}\label{eq:v_prof}
    v(\tau) = \left(\frac{v_{max}}{\sqrt{2}}\right) \sech{\left(\frac{\tau}{\tau_{0}}+\phi_0\right)},
\end{equation}
where $\phi_0 = \arcsinh(1)$.\footnote{The choice of any positive value $\phi_0$ gives Equation~\ref{eq:density} with a different normalization constant.} The constant $\tau_0$ is determined by setting $v(0) = v_{max}$ and $v(\tau_{d}) = v_{min}$, resulting in
\begin{equation}
    \tau_0 = \tau_{d} \left[\arcsech{\left(\frac{\sqrt{2}v_{min}}{v_{max}}\right)}-\phi_0\right]^{-1},
\end{equation}
where the ejection duration $\tau_{d}$ is set to $100~$ms \citep{Dessart2009, Perego2014, Kasen2017, Radice2018}, and $v_{min}$ and $v_{max}$ are the ejecta velocities at the inner and outer boundaries of the line-forming region, respectively. In this simulation study, they are fixed to 0.08~c and 0.35~c, respectively \citep{Hotokezaka2013, Siegel2017, Radice2018, Wollaeger2019, Rosswog2022}.
The velocity and its derivative profiles are presented in Figure~\ref{fig:velocity}. For a constant mass ejection rate, $\dot{\rm M}_{\rm ej}(\tau) = \dot{\rm M}_{\rm ej}$, the incorporation of this velocity profile into the continuity equation for stationary coasting shells provides (Eq. 38 in \citealt{uhm2011}; see also \citealt{Tak2023})
\begin{equation}\label{eq:density}
    \rho = \rho_0 v^{-3}t^{-3},
\end{equation}
where $\rho_0$ is the normalization constant, and $t$ is time since the explosion. This profile finds widespread adoption in kilonova studies \citep[e.g., ][]{Tanaka2017, Watson2019, Gillanders2022}. We assume a photospheric radius of $r_{ph} = 10^{15}$ cm from the center of the explosion (see \citealt{Drout2017} for the estimates of the photospheric radius of AT2017gfo) and its velocity of $v_{ph}$ = 0.16~c, which gives the observation time of $t = r_{ph}/v_{ph} \approx 2.4$ days after the explosion. The bolometric luminosity of the photosphere is assumed to be L$_{\rm ph} = 4\pi r_{ph}^2 \sigma_{\text{R}} T^4$, where $\sigma_{\text{R}}$ denotes the Stefan-Boltzmann constant. We emphasize that the velocity and density profiles are crucial factors influencing the resulting kilonova spectrum \citep{Tak2023}. Despite their significance, we adhere to employing the simple profile (Equation~\ref{eq:density}) throughout the entire simulation to focus on the impact of ejecta temperature and element mass.

Other than the aforementioned considerations, we adopt the configuration utilized by \cite{Tak2023}, including the selection of atomic dataset, plasma treatment methods, and abundance. The atomic dataset is complied from extensive sources including Chianti, Kurucz and DREAM, among others
\citep{Dere1997_Chianti, Biemont1999_DREAM, Kurucz2017, Kurucz2018, Quinet2020, Gillanders2021, Zanna2021, McCann2022, Bromley2023}
\citep[see][for details]{Tak2023}. For the \textsc{tardis} plasma setup, we adopt the special relativity treatment in radiative transfer \citep{Vogl2019}. We utilize the local thermal equilibrium ({\tt LTE}) approximation for ionization, while excitation follows the {\tt dilute-LTE} approach. Line interactions are addressed with the {\tt macroatom} approach \citep{Lucy2002, Lucy2003}, encompassing fluorescence effects and multiple internal line transitions. Lastly, we adopt the solar-system r-process composition, estimated by \cite{Prantzos2020}, for elements from Gallium (Z=31) to Uranium (Z=92). Among 49 heavy elements considered in this analysis, the mass fraction of 11 selected elements are listed in Table~\ref{tab:abud}. These elements are identified as the most active contributors to the kilonova spectrum in this simulation (see Section~\ref{sec:map}). We stress that our results are highly dependent on the relative abundance of elements (see Section~\ref{sec:ri}), and as a result, the strength of some features observed in AT2017gfo can be underestimated in this analysis, e.g., Ce\textsubscript{III} \citep{Domoto2021}.

In this study, we vary the ejecta mass in the line-forming region from $3 \times 10^{-6}$ M$_\odot$ to $0.3$ M$_\odot$ by changing the normalization constant $\rho_0$ and its temperature from 2000 K to 8400 K, while keeping other configurations fixed.\footnote{As in \cite{Gillanders2022} and \cite{Tak2023}, we assume a constant ejecta temperature.} These ranges are selected based on the results from hydrodynamic simulations and observations of AT2017gfo. The total ejecta mass is expected to be around 10$^{-4}$--10$^{-2}$ M$_\odot$ for the dynamical ejecta \citep[e.g.,][]{Bauswein2013, Hotokezaka2013, Abbott2017, Radice2018, Kullmann2022} and 10$^{-2}$--10$^{-1}$ M$_\odot$ for the disk-wind ejecta \citep[e.g.,][]{Perego2014, Siegel2017,Cowperthwaite2017, Villar2017}. Since the total ejecta mass of the system (M$_{\rm tot}$) is the sum of the mass beneath the photosphere (M$_{\rm ph}$) and the ejecta mass in the line-forming region (M$_{\rm ej}$), we adopt a wider mass range in this simulation. For the assumed ejecta velocity profile with the ejection time of 100~ms (Equation~\ref{eq:v_prof}), M$_{\rm ej}$ is approximately half of M$_{\rm tot}$ (M$_{\rm ej}$/M$_{\rm tot} \simeq 0.53$). The temperature of AT2017gfo decreases from 6000~K to 2500~K, from 1 to 5 days after the explosion \cite[see][]{Waxman2018}. 

Hereafter, `ejecta mass' refers specifically to the ejecta mass within the line-forming region, because this radiative-transfer simulation is not sensitive to the mass beneath the photosphere. 

\section{Results}\label{sec:result}

\subsection{Overall bolometric luminosity}
\begin{figure*}
  \centering
  \subfigure{\includegraphics[width=0.48\textwidth]{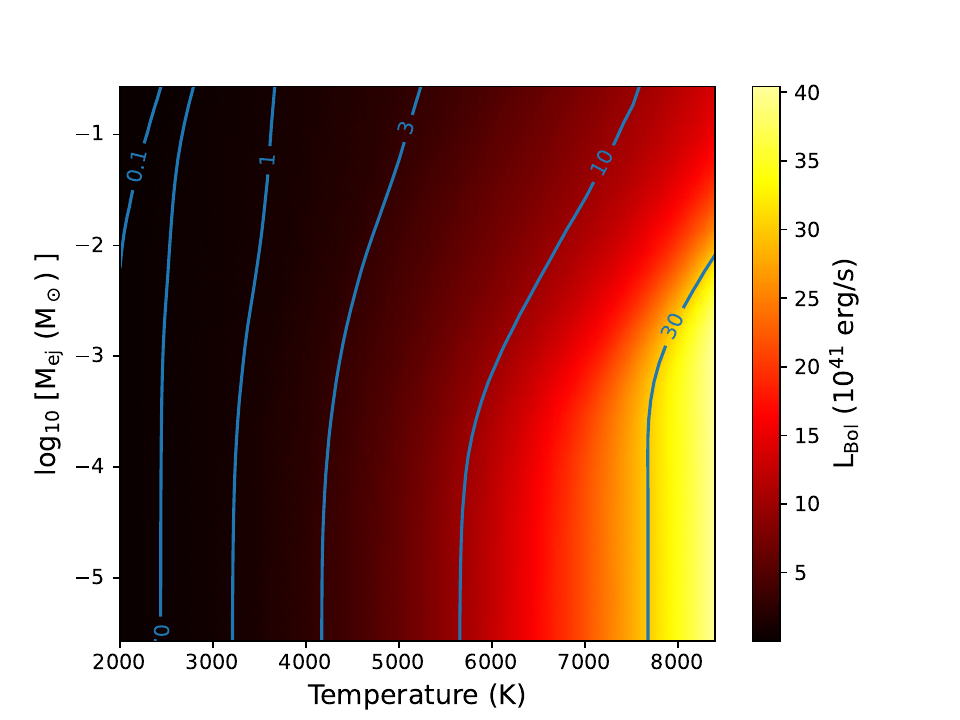}}
  \subfigure{\includegraphics[width=0.48\textwidth]{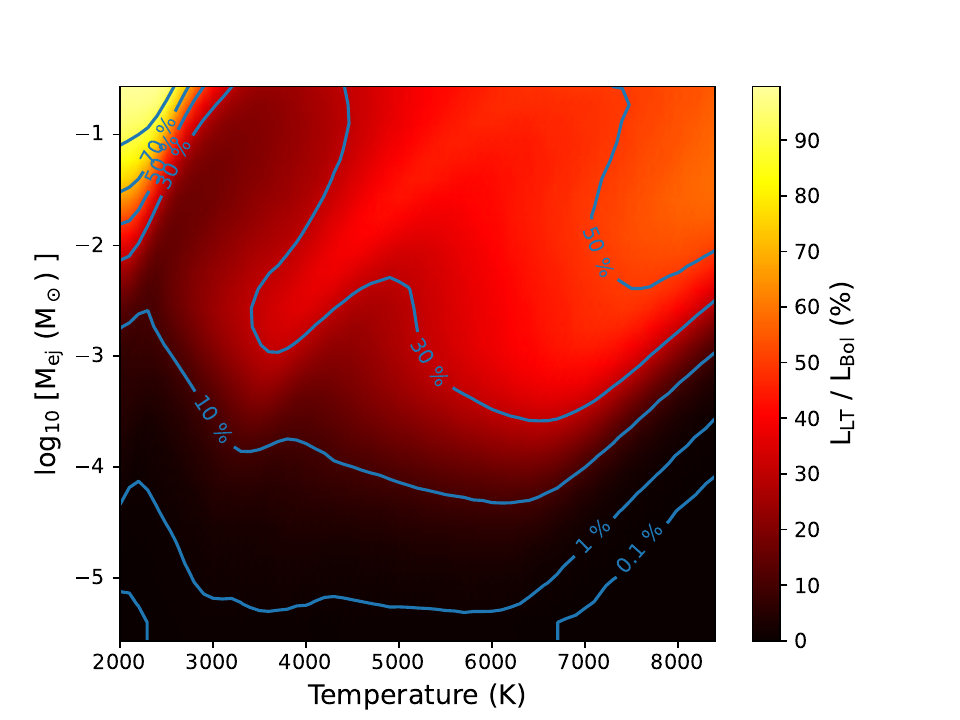}}
  \caption{Bolometric luminosity and the relative contribution from line transitions. The left panel displays the bolometric luminosity with contours (blue lines) observed in wavelengths ranging from 3000 \r{A} to 25000 \r{A}. The right panel shows the line-transition contribution, which represents the ratio of the bolometric luminosity of emitted photons interacting with heavy elements to the total bolometric luminosity. }
  \label{fig:tot_bol}
\end{figure*}

\begin{figure*}
  \centering
  \includegraphics[width=0.48\textwidth]{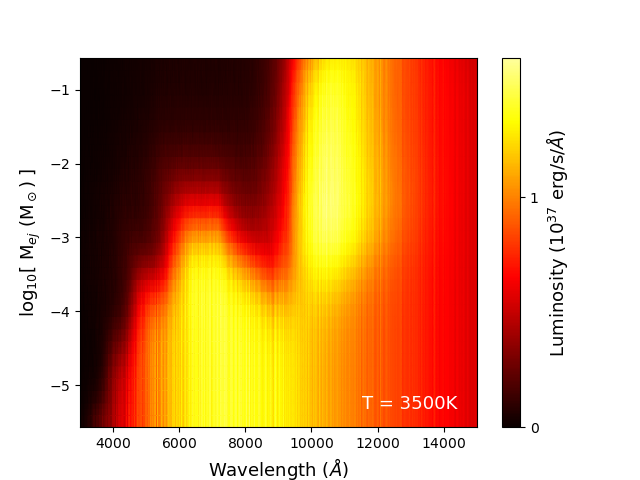}
  \includegraphics[width=0.48\textwidth]{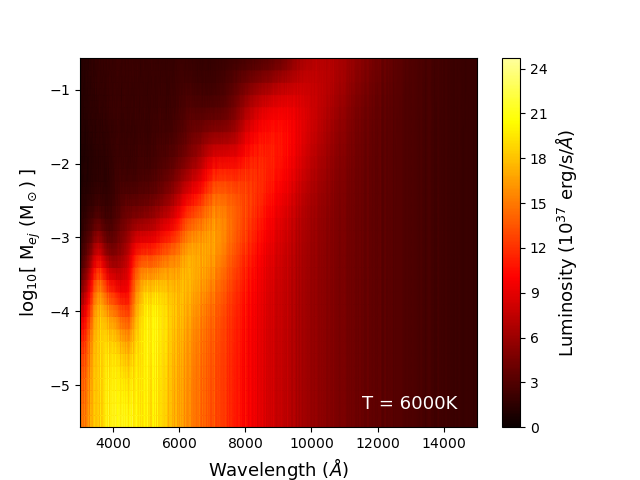} \\
  \caption{Spectral energy distributions  at temperatures 3500 K (left) and 6000K (right) for various ejecta mass in the line-forming region. Each spectrum obtained from a given ejecta mass is plotted along the x--z direction.}
  \label{fig:sed_in_two}
\end{figure*}

We synthesize more than 2000 kilonova spectra with various combinations of ejecta mass (M$_{ej}$) and temperature (\textit{T}), and calculate the bolometric luminosity (L$_{\rm Bol}$) across wavelengths ranging from 3000 \r{A} to 25000 \r{A} in the unit of erg/s. The overall contribution from line transitions for all the species under consideration is denoted as L$_{\rm LT}$. It is calculated as the sum of L$_{\rm Bol, s}$, L$_{\rm LT} = \sum_{\rm {\tiny Z=31}}^{\rm {\tiny Z=92}}$ L$_{\rm Bol, s}$, where L$_{\rm Bol, s}$ represents the contribution to the bolometric luminosity from each species. This value is derived by summing the luminosities of photon packets whose last interaction involves that particular species.

Figure~\ref{fig:tot_bol} shows the bolometric luminosity L$_{\rm Bol}$ (left panel) and the overall contribution of line transitions to the kilonova spectrum L$_{\rm LT}$/L$_{\rm Bol}$ (right panel). Since the photospheric bolometric luminosity L$_{\rm ph}$ increases with temperature, it is expected that the bolometric luminosity L$_{\rm Bol}$ will also increase with temperature. On the other hand, for a given temperature, an increase in M$_{\rm ej}$ reaches a point where L$_{\rm Bol}$ starts to decrease. This decrease occurs because some photon packets are scattered beneath the photosphere, and thus fail to escape the line-forming region. As evident in the right panel, the contribution of a line transition to the kilonova spectrum is inherently proportional to the ejecta mass. The contour displays a distinctive undulating pattern, particularly prominent in the 30\% and 50\% contours, indicating a significant dependence of the line-transition strength on the ejecta mass and temperature. This undulating feature is a result of the line-transition features of particular heavy elements (e.g., Sr\textsubscript{II}, Y\textsubscript{II}, and lanthanides) and their ionization states (see Section~\ref{sec:map}).

In Figure~\ref{fig:sed_in_two}, the stacked spectral energy distribution (SED) is depicted, with each SED plotted along the x--z direction. In the left panel, at a temperature of 3500 K, the brightest wavelength of the SED changes with increasing ejecta mass. When the ejecta mass surpasses 10$^{-3}$ M$_\odot$, emissions at wavelengths below 8000\r{A} tend to diminish due to numerous line transitions from various species, resulting in optically thick ejecta at these wavelengths. Instead, above this mass, a clear Sr\textsubscript{II} triplet signature is observed at the wavelength around 11000\r{A}, consistent with the observations of AT2017gfo \citep[see e.g.,][]{Watson2019, Domoto2021, Gillanders2022, Perego2022}. In the right panel, at a temperature of 6000 K, the luminosities at lower wavelengths ($<$ 8000\r{A}) are again diminished as the ejecta mass increases. This decrease is attributed to strong transitions of doubly ionized lanthanides (e.g., Pr\textsubscript{III} and Nd\textsubscript{III}; see Section~\ref{sec:map}).

\subsection{Luminosity fraction from individual elements}

\begin{figure*}
 \centering\includegraphics[width=0.95\textwidth]{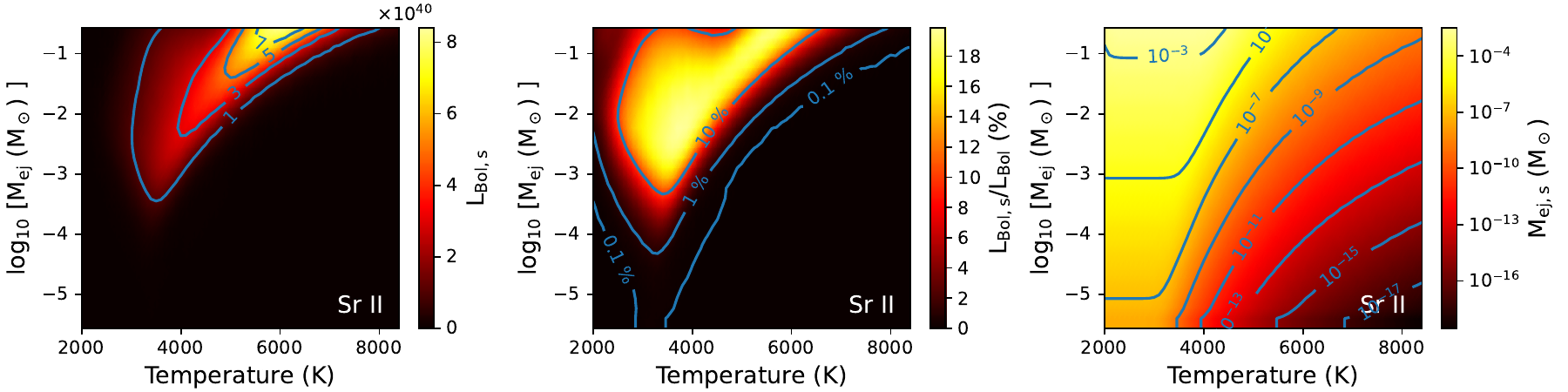}\\
 \includegraphics[width=0.95\textwidth]{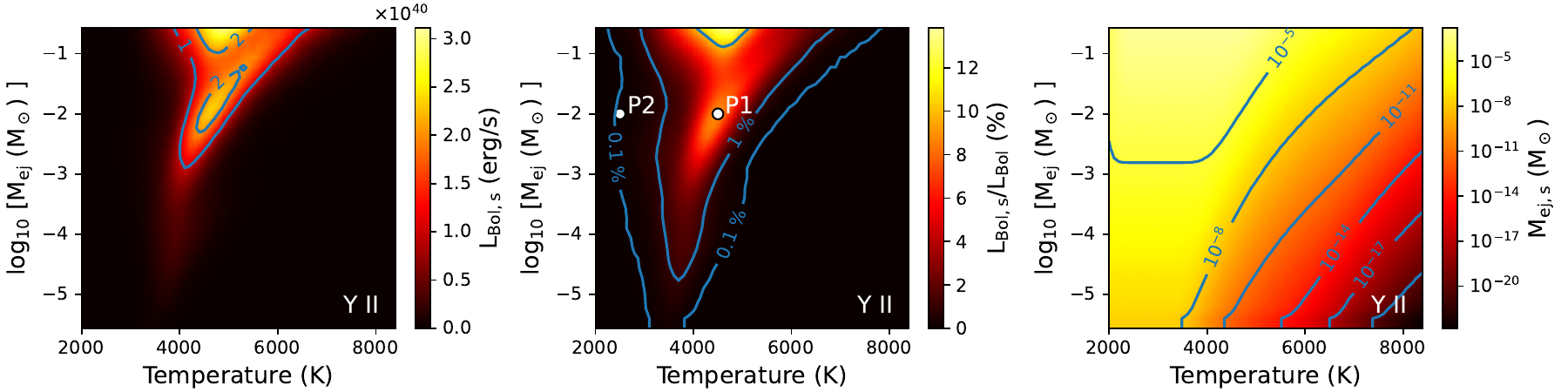} \\
 \includegraphics[width=0.95\textwidth]{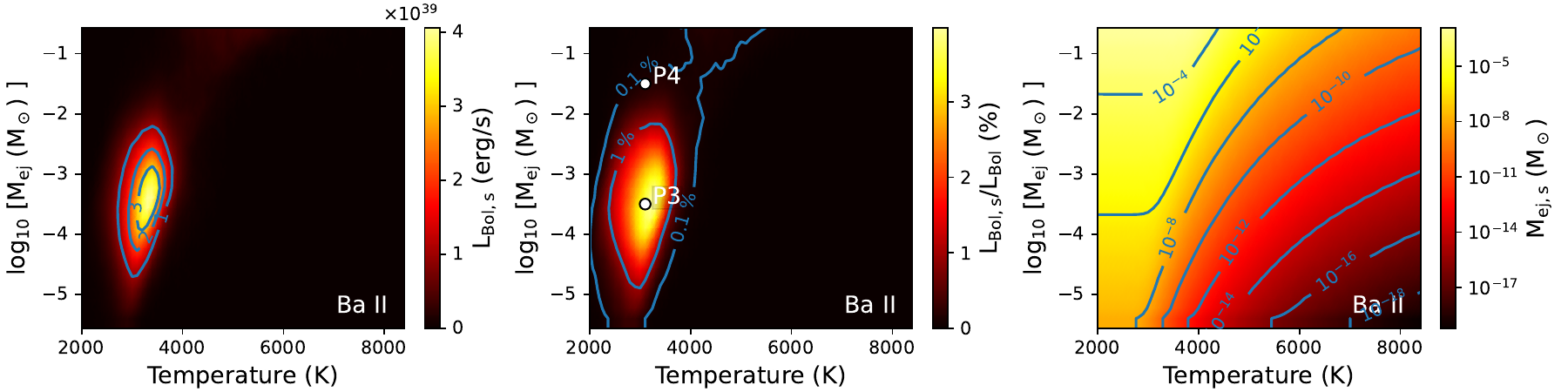} \\
 \includegraphics[width=0.95\textwidth]{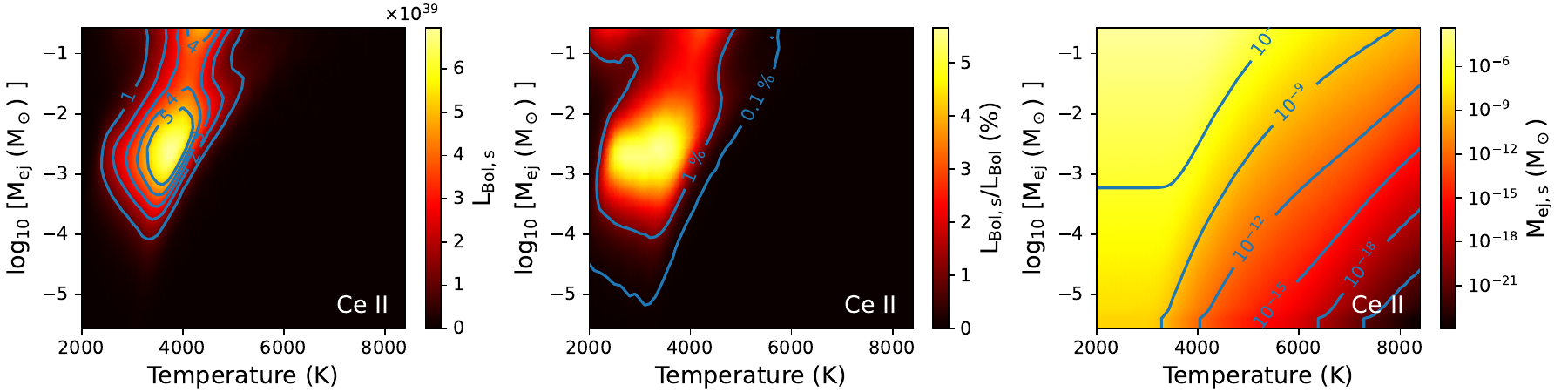} \\
 \caption{The luminosity and mass of four species (Sr\textsubscript{II}, Y\textsubscript{II}, Ba\textsubscript{II}, and Ce\textsubscript{II}) vary with temperature and ejecta mass. The left panels depict the bolometric luminosity of these four species, the middle panels illustrate their contribution to the overall bolometric luminosity, and the right panels display the mass of each species. Blue solid lines represent the contours. Points P1 to P4 in the middle panels of Y\textsubscript{II} and Ba\textsubscript{II} are used for studying the origin of the resonance island (Section~\ref{sec:ri}). }
 \label{fig:ri}
\end{figure*}

We examine the bolometric luminosity of each species (L$_{\rm Bol, s}$), its relative contribution to the total bolometric luminosity (L$_{\rm Bol, s}$/L$_{\rm Bol}$), and the mass of each species for a given ejecta mass and temperature. Here we caution that $L_{\rm Bol, s}$ does not correspond to some quantity of radiation produced by a given species. Instead, it is simply a derived quantity from our synthetic spectra that corresponds to the amount of emergent radiation from our \textsc{tardis} simulations that last interacted with a given species.

Figure~\ref{fig:ri} displays four examples -- Sr\textsubscript{II}, Y\textsubscript{II}, Ba\textsubscript{II}, and Ce\textsubscript{II} -- representing the most prominent species among considered heavy elements at temperatures below 6000 K, where some of them are prominently observed in AT2017gfo \citep[e.g.,][]{Watson2019, Domoto2021, Gillanders2022}. They contribute up to 3\% or more to the total bolometric luminosity and become resonant within specific parameter spaces of the ejecta mass and temperature. We call this region, the Resonance Island, and its location varies by species. Again, we clarify that this feature is not a result of the radiative decay of elements. Instead, it emerges from the reprocessing of photons as they pass through the line-forming region.

As depicted in the top-middle panel, Sr\textsubscript{II} exhibits the largest and strongest resonance island, making its signature easily detectable in the kilonova spectrum. The two species, Y\textsubscript{II} and Ce\textsubscript{II}, on the other hand, display multiple resonance islands; one for lower M$_{ej}$ and another for higher M$_{ej}$. The origin of the multiple islands is discussed in Section~\ref{sec:multi}. Species from Period 6 (Ba\textsubscript{II} and Ce\textsubscript{II}) typically exhibit resonance at lower M$_{ej}$ (approximately $10^{-3}$ M$_\odot$), while the resonance islands for Period 5 species (Sr\textsubscript{II} and Y\textsubscript{II}) are located at higher M$_{ej}$ and extends across a broad mass range. 

As evident in the right panels, the species' mass, M$_{ej, s}$, is directly proportional to M$_{ej}$. Simultaneously, an increase in temperature leads to a decrease in mass, which is attributed to the transition of a singly ionized species to a doubly ionized species, as described by the Saha ionization equation.\footnote{In thermal equilibrium, the ionization rate rises with temperature, as described by the Saha ionization equation: \mbox{$n_{i+1}/{n_i} \propto (k_B T)^{3/2} \exp{[-\epsilon_i/(k_B T)]}$}, where $n_{i+1}$ and $n_i$ represent the number of ions in the ($i+1)^{\rm th}$ and $i^{\rm th}$ ionization states, respectively, $\epsilon_i$ is the ionization potential between the ($i+1)^{\rm th}$ and $i^{\rm th}$ ionization states, and $k_B$ is the Boltzmann constant.}

The results for other elements can be accessed from the supplementary webpage.\footnote{\label{webpage}\url{http://kilonova.snu.ac.kr}}

\subsection{Resonance Island Map}\label{sec:map}
\begin{figure*}
  \centering
  \includegraphics[width=0.9\textwidth, trim={0 1.5cm 0 1.5cm}]{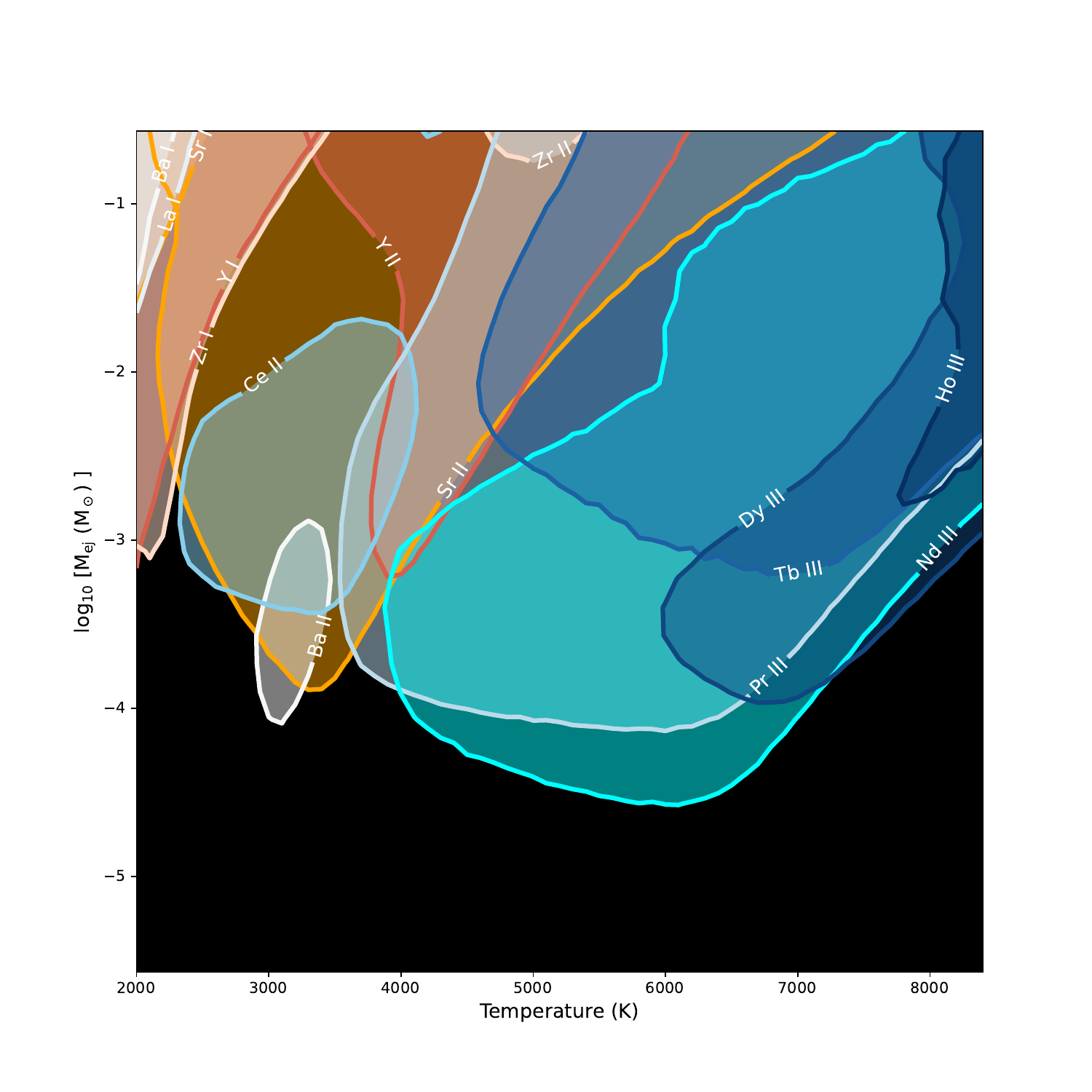}
  \caption{Resonance Island Map: The figure displays the 3\% contribution contours for 11 significant species. Neutral ions tend to contribute to the kilonova spectrum at lower temperatures. The resonance islands for singly ionized ions occupy the temperature range from 2500 K to 5000 K. Doubly ionized lanthanides contribute at higher temperature. Lanthanides are represented by bluish colors.}
  \label{fig:ri_map}
\end{figure*}

Figure~\ref{fig:ri_map} shows the resonance island map (RIM), where the 3\% contribution contours of the 11 most potent elements are displayed. As temperature decreases, the dominance of line transitions in the line-forming region shifts sequentially from doubly ionized to singly ionized and then to neutral elements. 

In ejecta with low temperature and high ejecta mass, prominent line transitions take place from neutral species (Sr\textsubscript{I}, Y\textsubscript{I}, Zr\textsubscript{I}, Ba\textsubscript{I}, and La\textsubscript{I}). The range between 2500 K and 5000 K is primarily dominated by line transitions from singly ionized species (Sr\textsubscript{II}, Y\textsubscript{II}, Zr\textsubscript{II}, Ba\textsubscript{II}, and Ce\textsubscript{II}), while the realm of doubly ionized species (lanthanides) extends above around 5000 K.

For singly-ionized atoms, Ba\textsubscript{II} and Ce\textsubscript{II} (Period 6 elements) tend to show their strong contributions to the kilonova spectrum at lower ejecta masses (around $10^{-3}$ M$_\odot$), while the resonance islands for Sr\textsubscript{II} and Y\textsubscript{II} (Period 5 elements) are situated at higher ejecta masses. 

This map clearly displays the resonance islands for various species, offering guidance to identify the relevant species associated with emission/absorption features in an observed spectrum, for a given ejecta mass and temperature.

\section{Discussion}\label{sec:disc}

\subsection{Origin of the resonance island}\label{sec:ri}

\begin{figure*}
  \centering
  \includegraphics[width=0.48\textwidth]{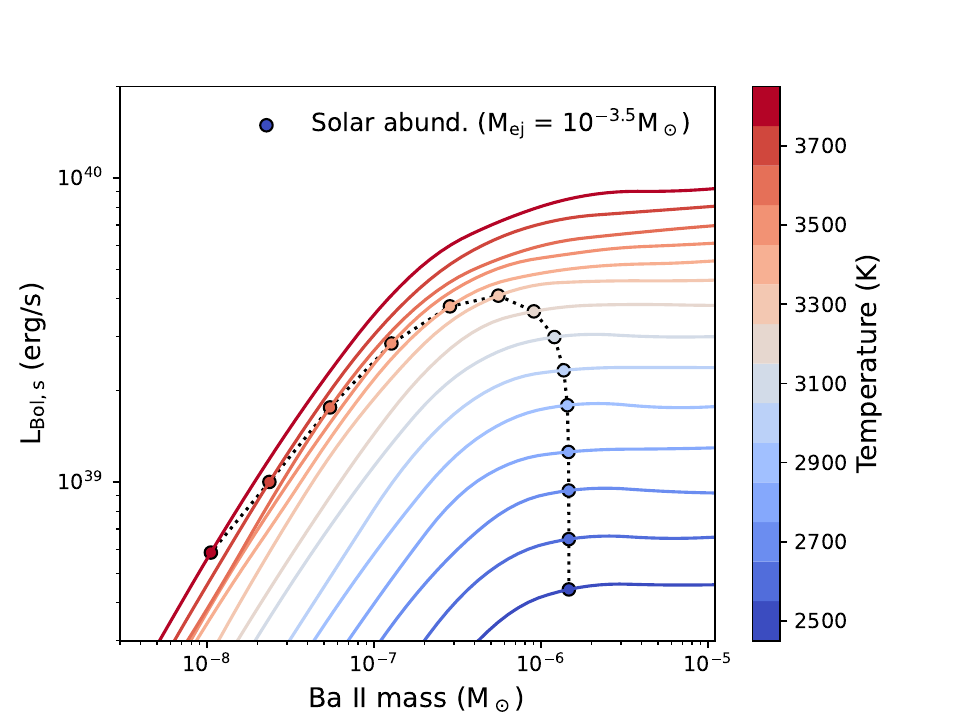}
  \includegraphics[width=0.48\textwidth]{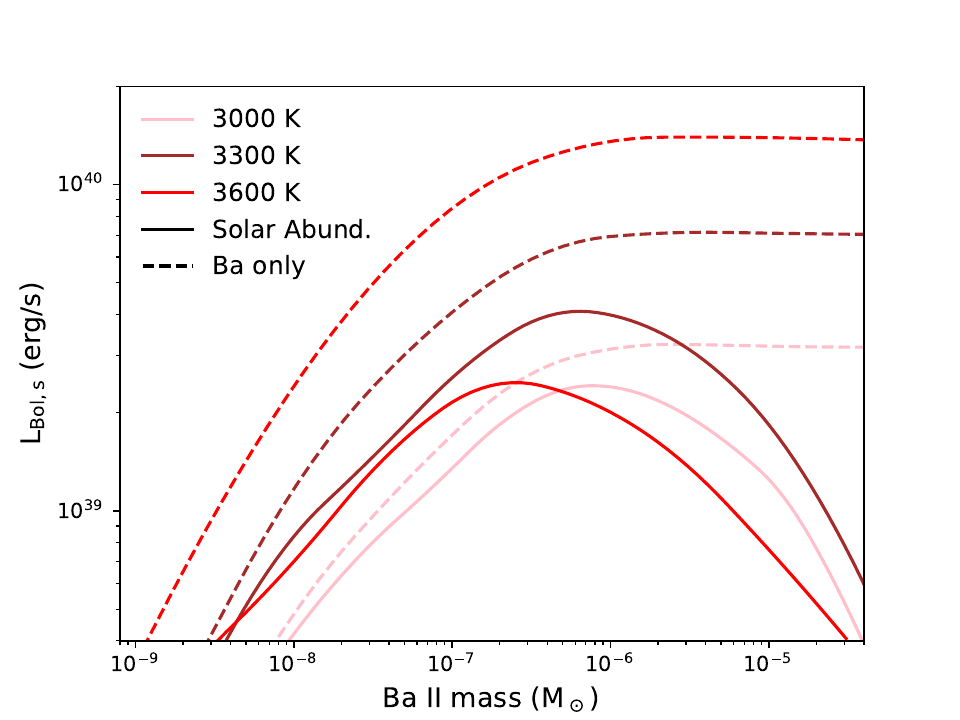} \\
  \caption{Bolometric luminosity of Ba\textsubscript{II} in various conditions. The left panel displays the bolometric luminosity of Ba\textsubscript{II} at different temperatures. Dotted points connected with the dashed line are obtained from the solar r-process abundance (Table~\ref{tab:abud}) for a fixed ejecta mass of M$_{\rm ej} = 10^{-3.5}$M$_\odot$. For each temperature, solid lines are obtained by varying the Ba mass while fixing the masses of others. The right panel illustrates the change in the bolometric luminosity of Ba\textsubscript{II} as we alter the ejecta mass for three temperatures. Solid lines correspond to results based on the solar r-process abundance, whereas dashed lines are derived under the assumption that only Ba is present in the line-forming region, meaning the ejecta mass is equivalent to the Ba mass.}
  \label{fig:mass_dep}
\end{figure*}

\begin{figure}
  \centering
  \includegraphics[width=0.48\textwidth]{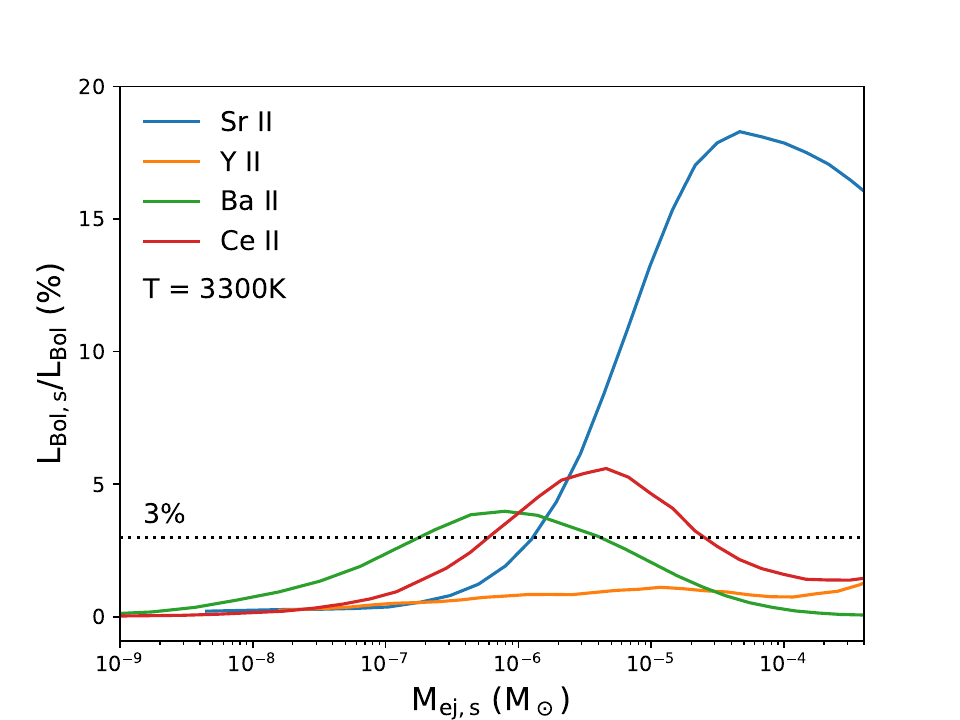} \\
  \caption{Relative contributions to the kilonova spectrum from four species at a temperature of 3300 K. The four species are Sr\textsubscript{II}, Y\textsubscript{II}, Ba\textsubscript{II}, and Ce\textsubscript{II}. }
  \label{fig:slice}
\end{figure}

When comparing the prominence of the resonance island (middle panels in Figure~\ref{fig:ri}) and the species mass at the island (right panels in Figure~\ref{fig:ri}), it becomes evident that the strength of resonance is not solely determined by the mass and/or temperature of the line-forming region. Given that Ba\textsubscript{II} has a well-isolated island, we focus on this region (specifically, the `P3' point in Figure~\ref{fig:ri}) to delve into the origin of the resonance island.

We begin by examining the origin for the horizontal boundaries of the resonance island for Ba\textsubscript{II}. The left panel of Figure~\ref{fig:mass_dep} shows how the bolometric luminosity of Ba\textsubscript{II} (L$_{\rm Bol, s}$) varies with the Ba\textsubscript{II} mass for a fixed ejecta mass (M$_{ej} = 10^{-3.5}$ M$_\odot$), represented by circle points connected with a dotted line. As the temperature increases from 2500 K to 3000 K, the mass of the species remains relatively constant, while L$_{\rm Bol, s}$ exhibits an increase. At around 3100 K, the ratio between singly and doubly ionized atoms starts to change, resulting in a decrease in Ba\textsubscript{II} mass. Nevertheless, L$_{\rm Bol, s}$ continues to rise until T $\simeq$ 3300 K, as the temperature-induced increase in L$_{\rm Bol, s}$ outweighs the reduction from the mass decrease. Subsequently, L$_{\rm Bol, s}$ declines above T $\simeq$ 3300 K due to the ionization. 

For a more detailed comprehension of the trend, we intentionally raise the Ba\textsubscript{II} mass in the line-forming region while keeping the masses of the other elements unchanged (represented by solid lines). Note that these artificial compositions no longer resemble the solar r-process abundance. For a fixed M$_{\rm ej, s}$, L$_{\rm Bol, s}$ continues to rise with increasing temperature, albeit at a diminishing rate. On the other hand, for a given temperature, L$_{\rm Bol, s}$ reaches a saturation point at around M$_{\rm ej, s} \sim 10^{-6}$ M$_\odot$. Beyond this point, additional mass increments no longer result in an increase in L$_{\rm Bol, s}$. This result implies that, even if there is more Ba\textsubscript{II} in the ejecta, we can only establish the lower limit of its mass. 

To be brief, the horizontal boundaries are linked to variations in the likelihood of interaction with changes in both photon flux and element mass. When the photon flux in the photospheric spectrum increases/decreases with temperature, there is a corresponding rise/decline in the interaction rate between photons and species. Likewise, an increase/decrease in the mass of Ba\textsubscript{II} results in a higher/lower interaction rate.

Next, we turn our attention to the examination of the vertical boundaries of the Ba\textsubscript{II} resonance island. In the right panel of Figure~\ref{fig:mass_dep}, we explore the change of L$_{\rm Bol, s}$ as M$_{\rm ej}$ increases at three temperatures (solid lines labeled as `Solar Abund.'): T = 3000 K (pink), 3300 K (brown), and 3600 K (red). Across all mass regimes, L$_{\rm Bol, s}$ at T = 3300 K surpasses those obtained at the other temperatures. We remind that the temperature of 3300 K corresponds to the summit of the island, where it reaches its maximum brightness (the left panel of Figure~\ref{fig:mass_dep}). In the low mass regime, an increase in Ba\textsubscript{II} mass results in a corresponding increase in luminosity. As the mass of the species increases, L$_{\rm Bol, s}$ peaks at a certain mass, dependent on temperature, and then gradually begins to decrease.

To investigate the cause of the evolution of L$_{\rm Bol, s}$ as a function of M$_{\rm ej, s}$, we examine the evolution of L$_{\rm Bol, s}$ under the same conditions but with Ba exclusively present in the line-forming region (depicted by the dashed lines labeled as `Ba only'). In general, L$_{\rm Bol, s}$ from `Ba only' is higher than that of `Solar Abund.'. Unlike the `Solar Abund.' case, L$_{\rm Bol, s}$ peaks at more and less the same  point around M$_{\rm ej, s} \sim 10^{-6}$ and remains constant beyond it. This suggests that the overall value of L$_{\rm Bol, s}$, as well as its peak and decline, originates from the influence of other elements. 

Considering the RIM for the given r-process solar abundance (Figure~\ref{fig:ri_map}), as the ejecta mass increases -- leading to a rise in the mass of all elements within the ejecta -- photons actively interact with other elements, specifically Ce\textsubscript{II} and Sr\textsubscript{II}. As a result, the interaction rate with Ba\textsubscript{II} decreases, resulting in the upper boundary of the island. Similarly, at higher temperatures, there is a greater chance for photon packets to interact with Sr\textsubscript{II} and Pr\textsubscript{III} (Figure~\ref{fig:ri_map}), which enlarges the gap in L$_{\rm Bol, s}$ between `Ba only' and `Solar Abund.'. This suggests that the relative abundance of elements in the ejecta ($Y_e$) can notably impact the resonance-island boundaries, particularly the upper limit. Furthermore, this result emphasizes the importance of the element composition in determining the contribution of each species, thereby shaping the kilonova spectrum. Further exploration of the relationship between $Y_e$ and the RIM can be explored in future research.

Finally, Figure~\ref{fig:slice} demonstrates how the relative contributions of Sr\textsubscript{II}, Y\textsubscript{II}, Ba\textsubscript{II}, and Ce\textsubscript{II} to the kilonova spectrum change as their masses vary at a temperature of 3300 K. The degree of influence between a species' mass and the strength of its island differs among species; e.g., the strength of Sr\textsubscript{II} resonance island increases sharply as its mass increases. In addition, the required masses of species for compounding the resonance island (3\% contour) differ from one another: specifically, Ba\textsubscript{II} requires a mass of $2\times10^{-7}$ M$_\odot$, Ce\textsubscript{II} needs $6\times10^{-7}$ M$_\odot$, Sr\textsubscript{II} demands $1.3\times10^{-6}$ M$_\odot$, and Y\textsubscript{II} forms their islands at its mass above $10^{-3}$ M$_\odot$ at this temperature. As described in Section~\ref{sec:map}, the resonance islands for Period 6 elements are formed at a lower ejecta mass compared to Period 5 elements. Given that the necessary mass for making the resonance island prominent varies among species, this pattern is likely due to the varying inherent strength, or probability, of species' line transitions. This highlights the influence of the distinct characteristics of each species' energy levels and corresponding line properties on the formation for the resonance island. 

In summary, the boundaries of the resonance island are determined by the combined effects of photon flux determined by temperature, species mass influenced by abundance and ejecta mass, line properties, and the relative line strength of other species. 

\subsection{Origin of the multiple resonances}\label{sec:multi}

\begin{figure}
  \centering
  \includegraphics[width=0.48\textwidth]{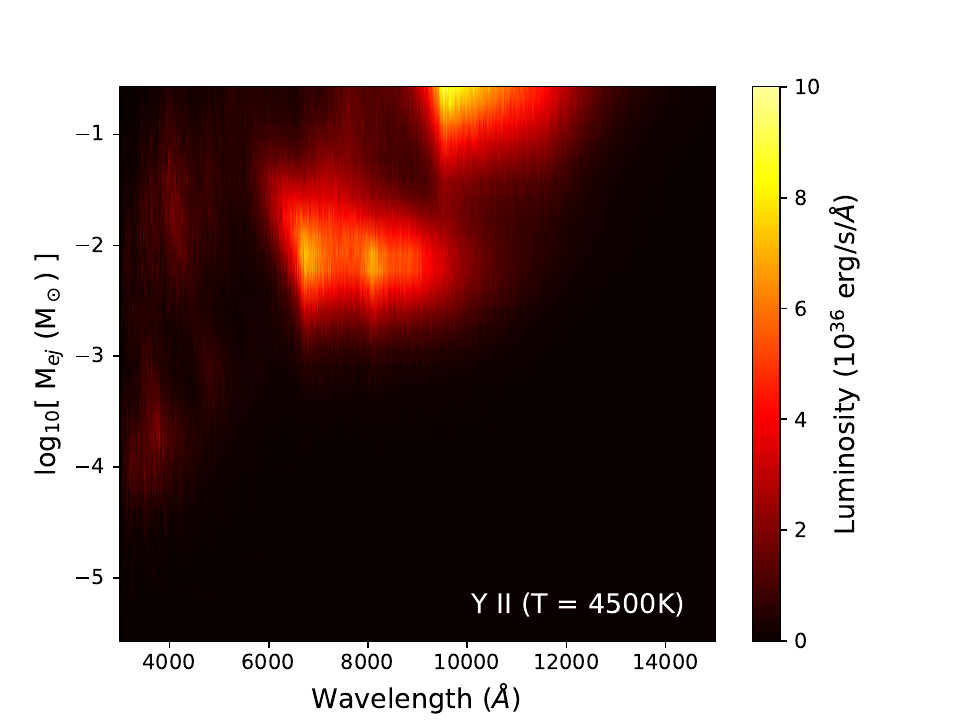}
  \caption{The contribution to the spectrum by Y\textsubscript{II} as a function of the ejecta mass. This spectral energy distribution is obtained at a temperature of 4500 K. The three bright regions correspond to the resonances of three transition lines.}
  \label{fig:YII}
\end{figure}

To understand the origin of the two distinct islands in Y\textsubscript{II}, we analyze changes in the spectral energy distribution (SED) with increasing ejecta mass (or species' mass).\footnote{The spectra for other elements at different temperatures are located at \url{http://kilonova.snu.ac.kr}.} For this investigation, we focus on a fixed temperature of 4500 K, as this temperature intersects both resonance islands. 

Figure~\ref{fig:YII} displays the stacked SED for the luminosity contribution by Y\textsubscript{II}. In the figure, three prominent points are visible; two are prominent at M$_{\rm ej} \sim 10^{-2}$ M$_\odot$ (approximately M$_{\rm ej, s} \sim 10^{-5}$ M$_\odot$) and the other is at a higher mass, M$_{\rm ej} \gtrsim 10^{-0.5}$ M$_\odot$. They are linked to a series of transitions, including three prominent line transitions with wavelengths of 7884~\r{A}, 9479~\r{A}, and 11181~\r{A}. Note that multiple weaker transitions contribute to those prominent points as well \citep[see][]{Sneppen2023a}. This implies that these two separate islands are associated with different line transitions. One of them at a lower ejecta mass largely originates from the two transitions (7884 \r{A} and 9479 \r{A}), which have similar line properties; they are similarly saturated as M$_{ej}$ increases. The other island is predominantly from the line transition with $\lambda =$ 11181 \r{A}. 

\subsection{Strong transitions}

\begin{figure}
  \centering
  \includegraphics[width=0.5\textwidth, trim={1.5cm 1.5cm 1.5cm 1.5cm}]{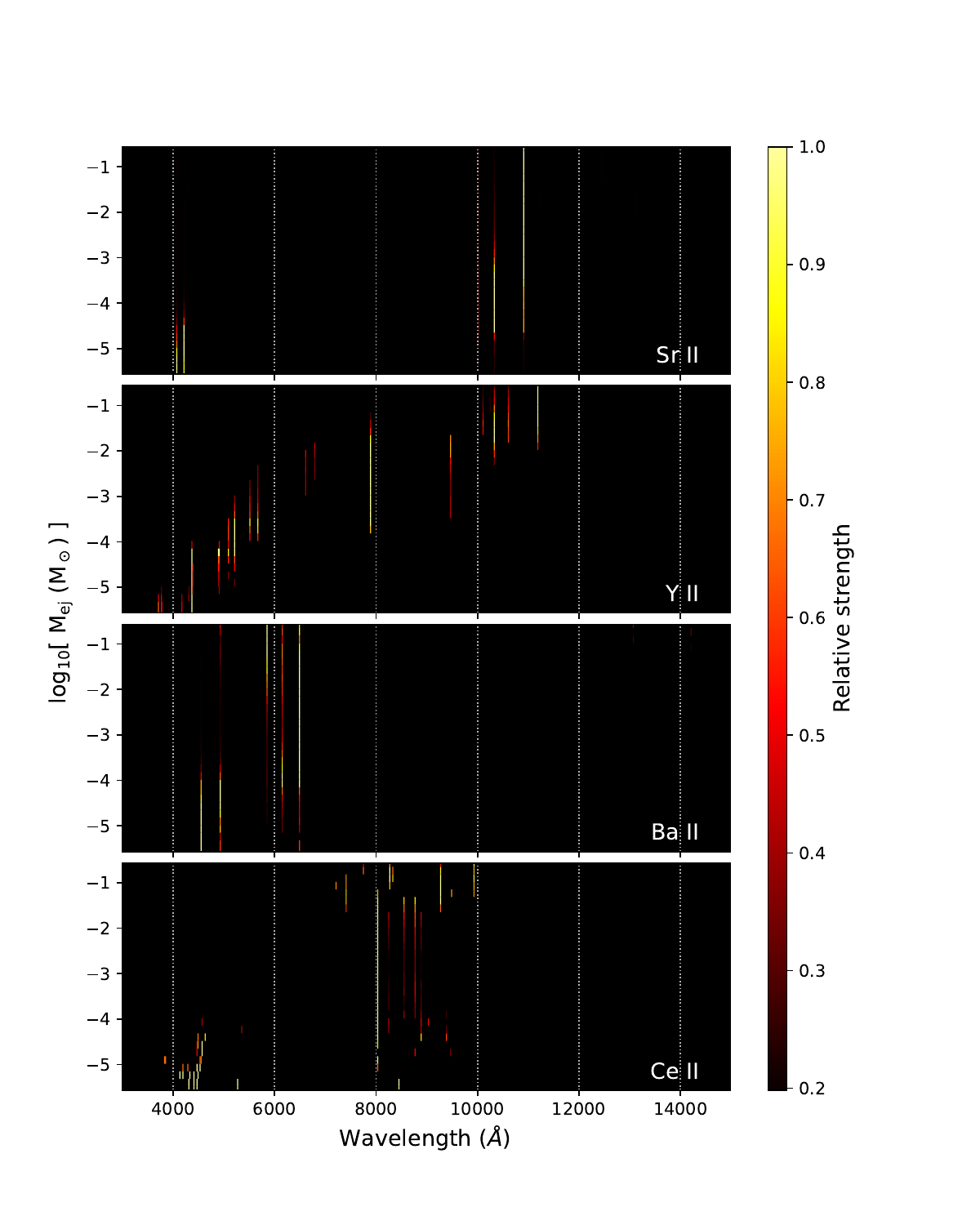}  
  \caption{The relative strength of line transitions of four species with respect to the ejecta mass at a temperature of 3500 K. We consider four representative species, Sr\textsubscript{II}, Y\textsubscript{II}, Ba\textsubscript{II}, and Ce\textsubscript{II}. The relative strength corresponds to the number of occurrences of the last line transitions for a given ejecta mass and temperature.}
  \label{fig:lines}
\end{figure}

The diagram in Figure~\ref{fig:lines} illustrates the strength of lines (i.e., the relative occurrence rate) for four elements (Sr\textsubscript{II}, Y\textsubscript{II}, Ba\textsubscript{II}, and Ce\textsubscript{II}) at different values of M$_{ej}$, all at a constant temperature of T = 3500 K. The results for other elements at different temperatures can be found on the supplementary webpage.\footref{webpage}

In the case of Sr\textsubscript{II} (top panel of Figure~\ref{fig:lines}), two prominent lines emerge at 10330 \r{A} and 10918 \r{A} from the NIR triplet transitions \citep[see e.g.,][]{Watson2019}. We note that the other triplet line (10039 \r{A}) is present in the models, but is fainter by a factor of $\sim 10$. Both Ba\textsubscript{II} and Ce\textsubscript{II} show a preferred line transition across a broad M$_{ej}$ range, with wavelengths of 6499 \r{A} and 8028 \r{A}, respectively. However, in the case of Y\textsubscript{II}, diverse and multiple transitions occur depending on M$_{ej}$ range, where the wavelength of strong transitions increases as M$_{ej}$ increases. This implies that the Y\textsubscript{II} feature (emission/absorption) in the observed spectrum would be smoothed out by its own other transitions, making it difficult to detect its signature \citep[this feature is also reported by][]{Sneppen2023a}. Note that the ejecta mass range of the preferred line domain (or vertical height) also changes with temperature.

\subsection{Observational implications}\label{sec:obs}

Observing the line-transition feature of a species necessitates meeting specific conditions, as the occurrence rate peaks at a particular temperature and is influenced by the species' mass. In other words, estimating the species' mass from observed spectra is only feasible when the conditions of temperature and ejecta mass are in harmony with those of the resonance island.

\begin{figure*} \centering\includegraphics[width=0.95\textwidth]{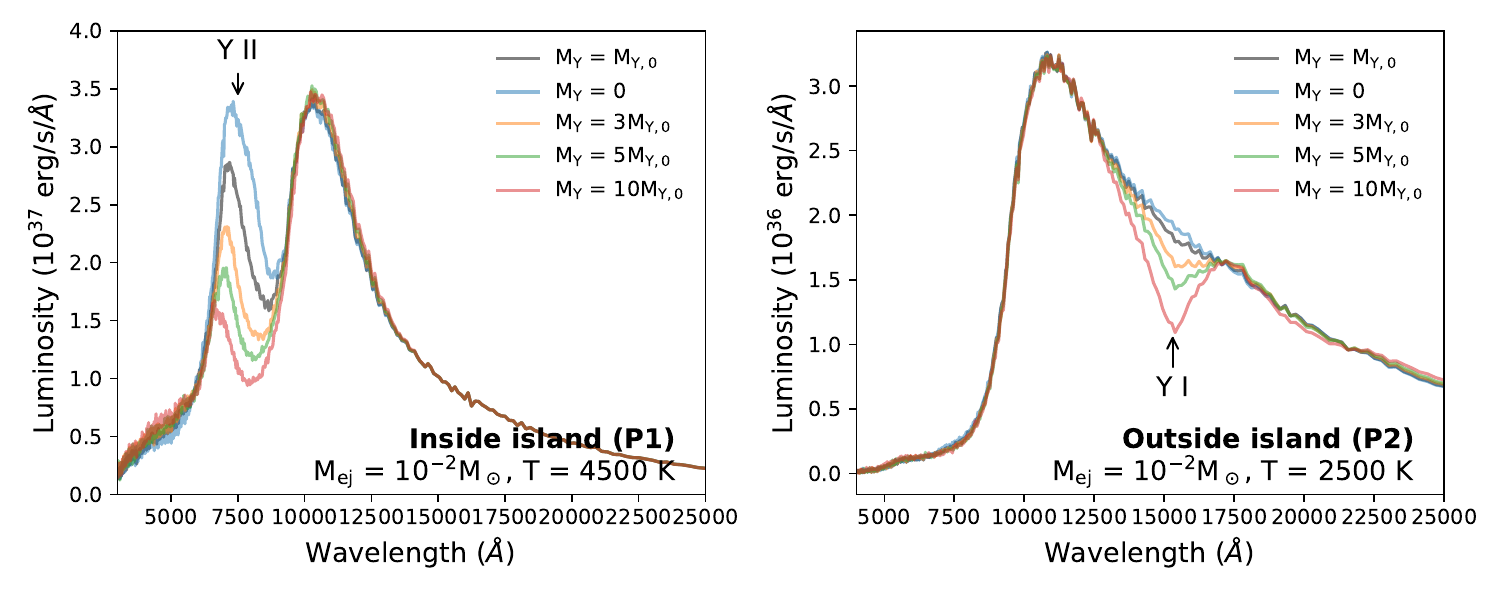}\\
 \includegraphics[width=0.95\textwidth]{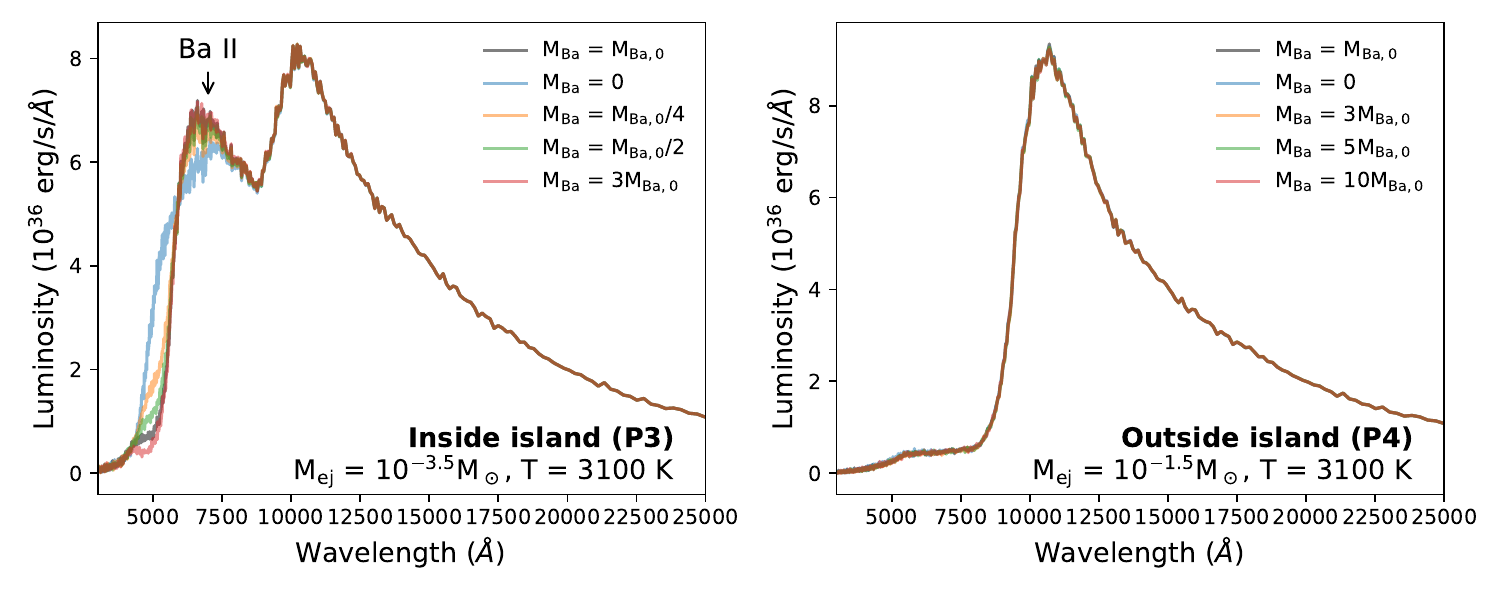} \\
 \caption{Spectral energy distributions under various conditions are depicted. The default conditions for each panel are displayed in the bottom-left corner, corresponding to P1 to P4 in Figure~\ref{fig:ri}. In each panel, we vary the mass of elements to observe how changes in their mass influence the resulting spectra. The left panels depict resonance conditions, wherein the mass of the target species (Y\textsubscript{II} or Ba\textsubscript{II}) notably influences the spectrum. The right panels correspond to nonresonance conditions. Note that, as observed in the top-right panel, the alteration in the mass of Y does not impact the contribution of Y\textsubscript{II} but does affect the contribution of Y I. }
 \label{fig:yeff}
\end{figure*}

To illustrate this point, we select two elements, Y\textsubscript{II} and Ba\textsubscript{II}, and investigate how observed SED can differ when the conditions agree/disagree with their respective resonance islands: in Figure~\ref{fig:ri}, P1 (Y\textsubscript{II} resonance island), P2 (outside of Y\textsubscript{II} resonance island), P3 (Ba\textsubscript{II} resonance island), and P4 (outside of Ba\textsubscript{II} resonance island). For each condition, we adjust the mass of a specific element, while keeping the masses of other elements fixed. In Figure~\ref{fig:yeff}, the top and bottom panels depict the changes in SEDs as we vary the mass of Y and Ba, respectively. When the ejecta mass and temperature align with resonance conditions (as shown in the left panels, P1 and P3 in Figure~\ref{fig:ri}), the changes in the masses of Y and Ba directly impact the SED in wavelengths of around 7500\r{A} and 6500\r{A}, respectively. For example, increasing the mass of Y reduces the emission at 7500 \r{A}, attributable to the increased opacity caused by Y\textsubscript{II} at that wavelength. This implies that it is possible to estimate the mass of the element from the spectral feature. Note that this Y\textsubscript{II} P-Cygni feature (at $\sim 7600$ \r{A}) was proposed as the source of an observed feature in the intermediate-phase spectra of AT2017gfo by \cite{Sneppen2023b}.

Under unfavorable conditions (nonresonance conditions), it might be challenging to identify any spectral features of the elements, even if they are present. In the lower-right panel, under conditions outside the Ba\textsubscript{II} resonance island (P4 in Figure~\ref{fig:ri}), it is evident that modifying the mass of Ba does not bring about any changes in the spectrum. In the case of Y (the top-right panel in Figure~\ref{fig:yeff}), altering the mass of Y does not induce the spectral feature of Y\textsubscript{II} in the wavelength around 7500 \r{A} as observed in the top-left panel. However, it does reveal the Y\textsubscript{I} absorption feature at around 15000 \r{A}, as the conditions of P2 are now aligned with those for the Y\textsubscript{I} resonance island (see Figure~\ref{fig:ri_map}).

\section{Summary}\label{sec:summary}
From the radiative-transfer simulations under various conditions, we identified the resonance conditions of the line transition(s) for each species. We also presented the RIM that serves as a guide for identifying element candidates that may exhibit spectral features as well as assessing the feasibility of constraining the mass of those elements at a given temperature.

Each species has its own resonance island with a unique shape. At a high temperature ($>$6000K), doubly ionized lanthanides contribute actively to the kilonova spectrum. The resonance island for neutral atoms is located at low temperatures with high mass ejection. The resonance-island boundaries are shaped by the combined impact of temperature, species mass, line transition characteristics, and the relative line strength of other elements. 

We have discussed several observational challenges in determining or constraining the mass of a species. In Section~\ref{sec:ri}, we highlighted that, independent of temperature, Ba\textsubscript{II} exhibits a mass saturation point in its strong line transition ($\lambda\sim$ 6500\r{A}) at M$_{\rm ej, Ba_{II}} \sim 10^{-6}$ M$_\odot$. Beyond this point, a further increase in the mass of Ba\textsubscript{II} does not impact the strength of the line feature. This implies that it becomes impractical to constrain the mass of Ba\textsubscript{II} if M$_{\rm ej, Ba_{II}} \gtrsim 10^{-6}$ M$_\odot$. We emphasize that this mass refers to the mass within the line-forming region, and the total mass of Ba\textsubscript{{II}} within the kilonova system in our parameter configuration will be approximately twice this value (assuming a uniform ejecta composition and a smooth, continuous density profile; see Section~\ref{sec:setup}). Consequently, when attempting to constrain the mass of a species, it is crucial to verify whether the determined mass is significantly lower than the saturation mass. Furthermore, when deviating from the resonance island for a species, the species would make a weak contribution to the observed spectrum, as discussed in Section~\ref{sec:obs}. These complexities pose challenges in straightforwardly estimating the mass of a species from the observed SED.

We stress that the accuracy of our result heavily depends on the completeness of the atomic dataset. Since there is limited information available for many of the heavy r-process elements, it is likely we underestimate the significance and strength of line transitions for certain heavy elements. Nevertheless, given the current understanding of the available atomic data, our result offers guidance for exploring the line signatures of kilonovae. As highlighted by \cite{Gillanders2023}, the properties of lines for certain species are inadequately examined, yet they hold particular importance in comprehending observations. This emphasizes the need for dedicated studies on the atomic data of those elements.

Given the current understanding of kilonova ejecta mass (M$_{ej} \sim 0.01-0.1$ M$_\odot$), elements associated with Period 5 are likely to exhibit strong line-transition features, as observed in AT2017gfo \citep[e.g.,][]{Watson2019, Domoto2021, Gillanders2022, Perego2022}. However, a future kilonova with lower mass ejections could exhibit prominent spectral features from Period 6 elements. In addition, rapid follow-up observations of kilonovae could allow us to identify a line-transition feature produced by doubly ionized lanthanides, since the ejecta temperature is higher at earlier times.

\acknowledgments
\section*{Acknowledgements}
We appreciate the kind assistance of Insu Paek from Seoul National University for hosting the supplementary website. This work was supported by the National Research Foundation of Korea (NRF) grant, No. 2021M3F7A1084525, funded by the Korea government (MSIT). This research made use of \textsc{tardis}, a community-developed software package for spectral synthesis in supernovae. The development of \textsc{tardis} received support from GitHub, the Google Summer of Code initiative, and from ESA's Summer of Code in Space program. \textsc{tardis} is a fiscally sponsored project of NumFOCUS. \textsc{tardis} makes extensive use of Astropy and Pyne.

\software{\textsc{tardis} \citep[][\url{https://zenodo.org/record/8128141}]{tardis}}

\bibliographystyle{aasjournal}
\bibliography{references}

\end{document}